\documentclass[conference]{IEEEtran}
\IEEEoverridecommandlockouts
\usepackage{cite}
\usepackage{amsmath,amssymb,amsfonts}
\usepackage{algorithm}
\usepackage{algorithmic}
\usepackage{graphicx}
\usepackage{graphics}
\usepackage{textcomp}

\usepackage{subcaption}
\usepackage{multirow}
\usepackage[inline]{enumitem}
\usepackage{gensymb}

\def\BibTeX{{\rm B\kern-.05em{\sc i\kern-.025em b}\kern-.08em
    T\kern-.1667em\lower.7ex\hbox{E}\kern-.125emX}}

\begin{document}

%%
%% The "title" command has an optional parameter,
%% allowing the author to define a "short title" to be used in page headers.
% 
% \title{\textbf{tubGEMM}: Sparsity-Effective Energy-Efficient Matrix-Multiply Units for Low-Precision DLAs}
% \title{\textbf{tubGEMM}: Sparsity-Effective Area-Power-Energy Efficient Temporal-Unary Matrix-Multiply Units}
\title{\textbf{tubGEMM}: Energy-Efficient and Sparsity-Effective Temporal-Unary-Binary Based Matrix Multiply Unit}

\author{\IEEEauthorblockN{Prabhu Vellaisamy\IEEEauthorrefmark{1}\IEEEauthorrefmark{2}, Harideep Nair\IEEEauthorrefmark{1}\IEEEauthorrefmark{2}, Joseph Finn\IEEEauthorrefmark{1}, Manav Trivedi\IEEEauthorrefmark{1}, Albert Chen\IEEEauthorrefmark{1}, Anna Li\IEEEauthorrefmark{1}}
\IEEEauthorblockN{Tsung-Han Lin\IEEEauthorrefmark{2}, Perry Wang\IEEEauthorrefmark{2}, Shawn Blanton\IEEEauthorrefmark{1}, and John Paul Shen\IEEEauthorrefmark{1}}
\IEEEauthorblockA{\textit{\IEEEauthorrefmark{1}ECE, Carnegie Mellon University, \{pvellais, hpnair, jafinn, mtrivedi, albertc1, ayli1, rblanton, jpshen\}@andrew.cmu.edu}}
\IEEEauthorblockA{\textit{\IEEEauthorrefmark{2}MediaTek USA Inc., \{prabhu.vellaisamy, harideep.nair, tsung-han.lin, perry.wang\}@mediatek.com}}
}

\maketitle
%%
%% The "author" command and its associated commands are used to define
%% the authors and their affiliations.
%% Of note is the shared affiliation of the first two authors, and the
%% "authornote" and "authornotemark" commands
%% used to denote shared contribution to the research.
% \author{Harideep Nair, Prabhu Vellaisamy, Santha Bhasuthkar and John Paul Shen}
% % \email{trovato@corporation.com}
% \affiliation{%
%   \institution{Electrical and Computer Engineering\\
%   Carnegie Mellon University}
% %   \streetaddress{P.O. Box 1212}
% %   \city{Dublin}
% %   \state{Ohio}
% %   \country{USA}
% %   \postcode{43017-6221}
% }t

%%
%% By default, the full list of authors will be used in the page
%% headers. Often, this list is too long, and will overlap
%% other information printed in the page headers. This command allows
%% the author to define a more concise list
%% of authors' names for this purpose.
% \renewcommand{\shortauthors}{Nair, et al.}

%%
%% The abstract is a short summary of the work to be presented in the
%% article.
\begin{abstract}
General Matrix Multiplication (GEMM) is a ubiquitous compute kernel in deep learning (DL).
To support energy-efficient edge-native processing, new GEMM hardware units have been proposed that operate on unary encoded bitstreams using much simpler hardware. Most unary approaches thus far focus on rate-based unary encoding of values and perform stochastic approximate computation. This work presents \textit{tubGEMM}, a novel matrix-multiply unit design that employs hybrid temporal-unary and binary \textit{(tub)} encoding and performs exact (not approximate) GEMM. It intrinsically exploits dynamic value sparsity to improve energy efficiency. Compared to the current best unary design uGEMM, tubGEMM significantly reduces area, power, and energy by 89\%, 87\%, and 50\% respectively. A tubGEMM design performing 128x128 matrix multiply on 8-bit integers, in commercial TSMC N5 (5nm) process node, consumes just 0.22 mm\textsuperscript{2} die area, 417.72 mW power, and 8.86 $\mu$J energy, assuming no sparsity. Typical sparsity in DL workloads (MobileNetv2, ResNet-50) reduces energy by more than 3x, and lowering precision to 4 and 2 bits further reduces it by 24x and 104x respectively.
% Lowering the precision to 4 bits and 2 bits further reduces the energy consumption by 14x and 40x respectively.
\end{abstract}

%%
%% The code below is generated by the tool at http://dl.acm.org/ccs.cfm.
%% Please copy and paste the code instead of the example below.
%%
% \begin{CCSXML}
% <ccs2012>
%   <concept>
%       <concept_id>10010583.10010633.10010655</concept_id>
%       <concept_desc>Hardware~Standard cell libraries</concept_desc>
%       <concept_significance>500</concept_significance>
%       </concept>
%   <concept>
%       <concept_id>10010520.10010521.10010542.10010294</concept_id>
%       <concept_desc>Computer systems organization~Neural networks</concept_desc>
%       <concept_significance>500</concept_significance>
%       </concept>
%  </ccs2012>
% \end{CCSXML}

% \ccsdesc[500]{Hardware~Standard cell libraries}
% \ccsdesc[500]{Computer systems organization~Neural networks}

%%
%% Keywords. The author(s) should pick words that accurately describe
%% the work being presented. Separate the keywords with commas.
%% A "teaser" image appears between the author and affiliation
%% information and the body of the document, and typically spans the
%% page.
% \begin{teaserfigure}
%   \includegraphics[width=\textwidth]{sampleteaser}
%   \caption{Seattle Mariners at Spring Training, 2010.}
%   \Description{Enjoying the baseball game from the third-base
%   seats. Ichiro Suzuki preparing to bat.}
%   \label{fig:teaser}
% \end{teaserfigure}

%%
%% This command processes the author and affiliation and title
%% information and builds the first part of the formatted document.

% \pagestyle{plain}

\section{Introduction \& Background}

General matrix multiplication (GEMM) is the primary compute kernel within the fully-connected and convolution layers of deep neural networks (DNNs). Though traditionally implemented as software libraries \cite{lawson1979basic},
% \cite{nugteren2018clblast,cuBLAS}
dedicated GEMM hardware units have been introduced more recently, specifically in deep learning accelerators (DLAs)
%such as Google TPUs (Tensor Processing Units) 
%with associated MXUs (matrix multiply units). 
%Nvidia introduced tensor cores \cite{markidis2018nvidia} capable of performing 4x4 matrix multiplication with half-precision floating-point operands in a single clock cycle. With the recent industry push towards edge-native devices, they introduced the Jetson Xavier NX module \cite{ditty2018nvidia}, consisting of 48 cores for less compute-intensive workloads.
%with some sacrifice on inference accuracy.   
where the GEMM compute kernel is directly implemented as a matrix multiply unit \cite{jouppi2017datacenter}. %The primary building blocks for MXUs are MAC units (typically organized in a 2D array) and determine the sizes of the (sub)matrices that can be processed in one compute cycle. The MXUs, in turn, constitute the primary compute fabric for DLAs. 
This work focuses on a novel GEMM microarchitecture and 
direct CMOS implementation of low precision (2-8 bits) matrix multiply units targeting edge-native DLAs. %targeting edge-native DNN processing, which requires significant improvements on area, power and energy efficiency.

%GEMM operations are scaled MAC computations for matrix multiplications and are generally incorporated in the fully connected and convolution layers of DNNs. Though traditionally, GEMM is used as software libraries [][], recent works have demonstrated there is considerable potential in implementing dedicated GEMM units for hardware acceleration resulting in improved energy efficiency [][]. NVidia introduced tensor cores [] capable of performing 4x4 matrix multiplication with half-precision floating-point operands in just a single clock cycle. Recognizing the industry trend toward edge-native devices, they further introduced the Jetson Xavier NX module [], consisting of 48 cores for less compute-intensive workloads, albeit by sacrificing inference accuracy.   

%Conventional matrix multiply units are implemented for multi-bit binary encoded values. Recent research has suggested using unary encoding of values to achieve hardware efficiency, especially for low-precision computation \cite{ugemm}. Unary encoding expands the binary format into a serial bitstream format. Operations are perform on two bitstrems serially using very simple hardware. 
%utilize binary computing where computations and representations of  are in multiple parallel bits, and they suffer from poor bit-width scaling and exponential wire congestion growth [], resulting in poor hardware metrics. To combat this,

Lowering precision is fast becoming the de facto approach to increasing performance and decreasing energy consumption for DLAs. 16-bit formats for training are now widely used in industry. 16-bit accelerators are shown to provide 4x to 8x performance relative to 32-bit designs \cite{NEURIPS2020_13b91943}. Further, authors in \cite{NEURIPS2020_13b91943} report successful 4-bit based training on a collection of Deep Learning (DL) workloads with minimal accuracy loss.
% while providing significant hardware acceleration of 7x over state-of-the-art FP16 systems.
IBM researchers achieved 8-bit precision for training and 4-bit precision for inference across many DL datasets \cite{wang20188}.
Ongoing research demonstrates the efficacy of pursuing low-precision compute to boost the hardware performance of current DLAs, with the potential of moving down to 4-bit and 2-bit compute while maintaining close to ideal inference accuracy. 

\begin{figure}[t]
\centering
\includegraphics[width=0.9\columnwidth, height=3.3cm, keepaspectratio]{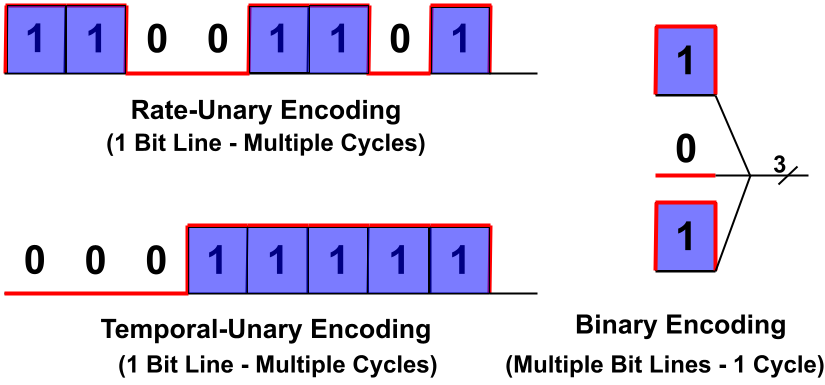}
\caption{Rate-Unary Encoding vs. Temporal-Unary Encoding vs. 3-Bit Binary Encoding for the value '5'}
\label{fig:enc}
\end{figure}

\textit{Unary computing} has been touted as a promising solution to building area and energy-optimized GEMM units \cite{gaines1967stochastic}. This paradigm substitutes the need for multiple parallel bits in binary computing with serial unary bitstreams, resulting in significantly simpler and more energy-efficient hardware \cite{alaghi2013survey, lee2018architecture, sim2017new, ugemm}. Unary computing can manifest in two forms of encoding: \textit{rate-based unary} and \textit{temporal-based unary}; and two formats: \textit{unipolar} (unsigned) or \textit{bipolar} (signed) \cite{ugemm}. Data representation in rate-based unary coding is based on the frequency of 1s and 0s within a bitstream, with the number of 1s in the bitstream proportional to the data value. For temporal-based unary encoding, data is represented in the form of a sequence of 1s followed by a sequence of 0s, with the number of consecutive 1s representing the value. In contrast to temporal coding, rate coding results in approximate computation and suffers from the correlation problem \cite{baker2019impact}. Fig. \ref{fig:enc} illustrates the three data encoding schemes of: 1) rate-based unary, 2) temporal-based unary, and 3) conventional binary.
%Existing literature in unary computing focuses mainly on \textit{rate}-based \textit{stochastic} computing, trading off accuracy for hardware-friendly implementation. Rate coding results in approximate computation as it suffers from the correlation problem \cite{baker2019impact}.

Recent work on \textit{unified}-unary GEMM design, uGEMM \cite{ugemm}, depicts the state-of-the-art in all key hardware metrics. 
%However, no existing work so far has fully incorporated unary \textit{temporal}-encoding with its resultant energy efficiency. 
Although it supports temporal encoding, the underlying hardware is still rate-based targeting stochastic computing. 
While uGEMM focuses on \textit{unified} encoding-agnostic hardware, our goal is to extract maximum area-power-energy efficiency by specializing the hardware for temporal encoding. 
%Moreover, the high-accuracy and low latency results for uGEMM are shown for rate-encoded bitstreams; temporal bitstreams incur significant accuracy loss to achieve uGEMM's reported energy efficiency. 
Further, we believe, GEMM hardware targeting current DLAs should ideally perform \textit{exact} compute with no accuracy loss.
%, unlike current unary GEMM hardware for stochastic computing.

In this work, we present \textit{tubGEMM}, a highly efficient GEMM unit operating on temporal-unary and binary (tub) encoded values and using temporal processing hardware design, targeting deployment in low-precision edge-native DLAs.
% based on a new temporal algebra \cite{smith2018space}. 
%This approach is ideal for resource-constrained edge devices. 
tubGEMM performs \textit{exact} compute and does not incur accuracy loss like current unary stochastic GEMM hardware. It also utilizes a modified temporal encoding, termed \textit{twos-unary}, to optimize compute latency with minimal hardware overhead.
% This paper focuses on the microarchitecture design of tubGEMM as a matrix multiply unit.

The major contributions of this paper are as follows:
% \vspace{-2pt}
\begin{itemize}
\setlength\itemsep{0.1em}
    \item We present a novel GEMM design, \textit{tubGEMM}, employing hybrid \textit{temporal-unary and binary} ``tub'' approach with a modified \textit{twos-unary} temporal encoding.
    % that significantly improves area, power, and energy relative to current unary designs.
    \item Unlike previous unary GEMM approaches that operate on \textit{rate-unary} encoded bitstreams for approximate computing, tubGEMM performs exact computing.
    \item tubGEMM is 9x, 8x and 2x more efficient in area, power, and energy (i.e., achieves 89\%, 87\%, and 50\% reduction) respectively than state-of-the-art unary design \textit{uGEMM}, while incurring no inference accuracy loss. 
    \item We perform rigorous industry-standard evaluation of tubGEMM's hardware complexity scaling using commercial design tools and the current TSMC N5 (5nm) process node. We scale the input matrix dimensions from 16x16 to 128x128 and precision from 2 bits to 8 bits, and report post-synthesis power-performance-area (PPA) results.
    \item We demonstrate that tubGEMM can naturally exploit data value sparsity on actual workloads (MobileNetv2 and ResNet-50) to dynamically improve its effective compute latency, power consumption and energy consumption. 
    % \item A quick assessment of memory bandwidth shows that the tubGEMM designs have the potential to significantly alleviate the memory bandwidth bottleneck. 
\end{itemize}

\section{Value Encoding and Processing}
Traditional multiply-accumulate (MAC) units operate on binary-encoded inputs (Fig. \ref{fig:enc}) using bit-parallel multipliers and adders that deliver low latency at the cost of high area and power.
% The low latency benefits of such designs can be outweighed by high area and power costs.
%previously "the low latency benefits are often outweighed". This sort of implies that we shouldn't use binary, but that's clearly not the case. I can't think of a great way of wording it though
% Good point
%by high area and power costs due to spatial binary representation, especially in edge computing scenarios where the hardware must operate within highly limited area and power budgets.
Unary encoding with bit-serial processing is a viable alternative that tries to trade off latency for improved area and power. As shown in Fig. \ref{fig:enc}, it encodes values either stochastically in the frequency of randomly distributed ones within a bitstream (rate-unary encoding) or deterministically in the number of consecutive ones in a bitstream (temporal-unary encoding). Our work focuses on temporal-unary encoding in contrast to rate-unary encoding for three reasons: 1) it avoids costly random number generators, 2) it is capable of deterministic exact computation, and 3) it incurs lower dynamic power due to fewer edge transitions (only two transitions). 

Analogous to Smith's temporal algebra \cite{smith2018space}, 
%and the spike encoding and processing employed in neuromorphic temporal neural networks (TNNs) \cite{nair2021online},
temporal-unary encoding here uses an $n$-cycle wide logic pulse to represent a value $n$, and can thereby consume up to a maximum of $(2^b - 1)$ cycles to represent a $b$-bit value and $(2^b - 1)^2$ cycles to represent multiplication result of two such $b$-bit values. To reduce this latency while still reaping the benefits of temporal-unary encoding, we use a hybrid \textit{temporal-unary and binary} (tub) approach, similar to \cite{diwu2022uSystolic}. With only one of the inputs temporal-unary encoded and the other input kept in binary, the maximum multiplication latency is reduced from $(2^b - 1)^2$ cycles to $(2^b - 1)$ cycles. We further optimize this latency via a modified encoding as described in the next section.
% Fig. \ref{tubmac} illustrates the resultant MAC design, \textit{tubMAC}, that has added hardware to dynamically convert the \textbf{A} matrix value from conventional binary to temporal-unary format (Fig. \ref{convmac}). The other input values from matrices \textbf{B} and \textbf{C} remain in conventional binary format. Such tubMAC units with in-built encoders can directly replace conventional MAC units with binary interface and are the key building blocks in our tubGEMM design.
% It is to be noted that, in an array of tubMAC-based Processing Elements (PEs), the temporal-unary encoder can be shared across multiple tubMAC units within the PE array.
% 
%The added hardware for this conversion is accounted for in all the PPA analysis. 
%The PPA analysis in Section \ref{eval_sec} accounts for the hardware required to convert $\textbf{A}$ from binary to temporal-unary signal
%, and hence the proposed tubGEMM can be easily plugged into a binary interface, just like in \cite{diwu2022uSystolic}.

\begin{figure}[t]
\centering
\includegraphics[width=0.9\columnwidth]{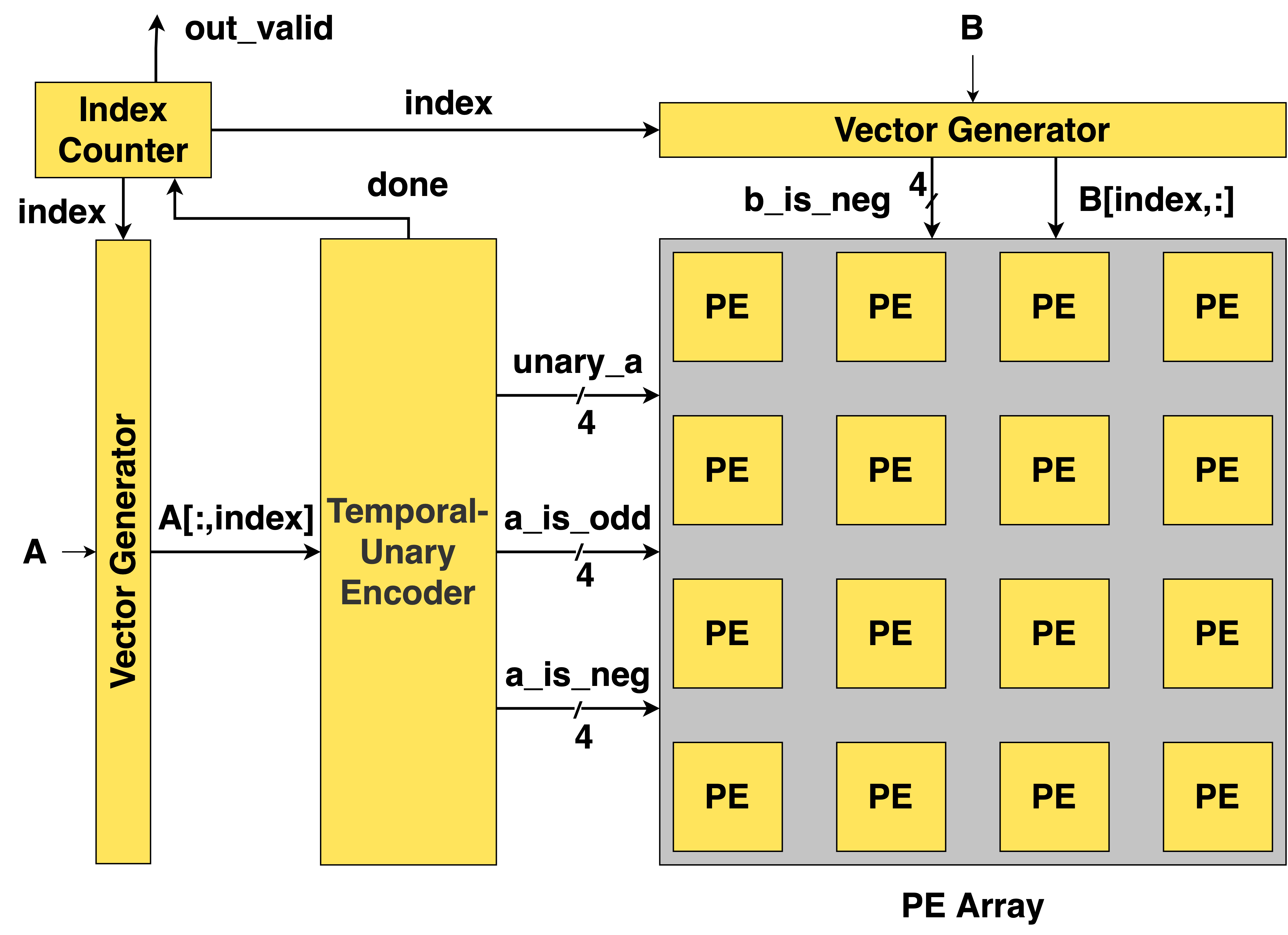}
\caption{4x4 tubGEMM Architectural Block Diagram}
\label{fig:tmxu}
\end{figure}

\section{tubGEMM Microarchitecture}
% HARI TODO: THIS SECTION JUST CONTAINS SKELETAL CONTENT. NEED EXTENSIVE MAKEOVER
% This section describes the tubGEMM microarchitecture and the dataflow used in detail.
In this section, we present our proposed \textit{tubGEMM} microarchitecture (Fig. \ref{fig:tmxu}) for matrix-multiply unit design, which consists of an $M$x$P$ array of processing elements (PEs) designed to perform: \textbf{Y} = \textbf{A} x \textbf{B} + \textbf{C}, where $\textbf{A}$, $\textbf{B}$, \textbf{C} are generic $M$x$N$, $N$x$P$ and $M$x$P$ input matrices, and $\textbf{Y}$ is the $M$x$P$ output matrix.
% This can be easily extended to non-scaled and scaled GEMM.
%, as will be described later.
% We also perform a qualitative gate-count evaluation for each of the components, thereby allowing us to assess the hardware complexity scaling with respect to number of dimensions and bit width. 
Alongside the tubGEMM microarchitecture, we propose a modified temporal-unary encoding scheme, \textit{Twos-Unary (2-Unary)}, that significantly reduces the latency with minimal hardware overhead. This novel encoding optimization is described in the following subsection.

\subsection{Twos-Unary Temporal-Unary Encoding}
As discussed earlier,
% traditional temporal encoding incurs up to $(2^b - 1)$ cycles to encode a unary value, which results in
the hybrid \textit{tub} approach incurs upto $(2^b - 1)$ cycles
% consuming the same number of cycles
for a single multiplication, assuming each unary bit represents magnitude of 1. 
%We refer to this as \textit{1-Unary}. 
The encoding latency can be halved if each unary bit represents a magnitude of 2 instead. The only overhead required is a correction mechanism for odd values. This is done with no extra hardware as the existing adder can be reused to add one additional value in the last cycle (described further in the following subsections). With the \textit{2-Unary} temporal encoding, the maximum latency is only $(2^{b-1})$ cycles, half of the original $(2^b - 1)$ cycle latency.
%, accounting for the additional cycle overhead for odd number correction.

This encoding can be generalized to arbitrary $n$ (\textit{n-Unary}), where $n$ is a power of 2. 
%Note that non-powers of 2 are unreasonable since they require additional adders and shifters in the hardware to account for the modulo-$n$ correction. 
However, with higher powers of 2 (e.g., 4, 8), the correction mechanism requires an increasing number of shifters. Our exploration found that $n$=2 provides a good balance between additional complexity and latency reduction. Next, we describe how the input matrices are streamed into tubGEMM, utilizing the \textit{2-Unary tub} approach.

\subsection{Input Matrices Dataflow}

%\begin{algorithm}[t]
%\caption{tubGEMM Matrix Multiplication Dataflow}
%\label{alg:cap}
%\begin{algorithmic}
% \Require $n \geq 0$
%\ENSURE $\textbf{Y} = \textbf{A} * \textbf{B}$
%\STATE $Y \gets 0$
%\FOR{$k = 0:N-1$}
% \COMMENT{One computation step updates all nodes based on $k^{th}$ column of A and $k^{th}$ row of B in parallel}
%\FOR{$i = 0:M-1$}
%\FOR{$j = 0:P-1$}
%\STATE $Y[i][j] \gets Y[i][j] + A[i][k] \times B[k][j]$
% \COMMENT{Can take upto $2^{b-2}$ cycles}
%\ENDFOR
%\ENDFOR
%\ENDFOR
%\end{algorithmic}
%\end{algorithm}

Fig. \ref{fig:tmxu} shows the block diagram for the proposed tubGEMM with $M$x$P$ nodes in the PE array. It takes in temporal-unary \textbf{A} input from the left (after the binary to the temporal-unary encoder) and binary \textbf{B} input from the top.
% The key dataflow mechanism is summarized in Algorithm 1 and described next. 
Input \textbf{A} arrives one column ($M$ elements) at a time, and \textbf{B} arrives one row ($P$ elements) at a time, in lockstep fashion. Both row and column are indexed identically within their corresponding matrices, i.e., $k^{th}$ column of \textbf{A} arrives with $k^{th}$ row of \textbf{B}, and their outer product is computed. Thus, the computation occurs in $N$ \textit{steps} where each \textit{step} performs a single \textit{column-row} outer product (\textbf{A} has $N$ columns, and \textbf{B} has $N$ rows).
% At each \textit{step}, the $M$x$P$ PEs are updated with the $M$x$P$ output values, taking as many cycles as the magnitude of the maximum temporal-unary input value.
The $M$x$P$ PEs keep accumulating the $M$x$P$ output values from the outer product after each \textit{step}, taking as many cycles as the magnitude of the maximum temporal-unary input value in every \textit{step}.
If we initialize the $M$x$P$ PEs with \textbf{C} matrix values, it computes \textbf{Y} at the end of $N$ \textit{steps}. 
% A \textit{step\_done} signal is generated after every $N$ \textit{steps}. 
Other than the PE array (described in Section \ref{sec:mac}), the proposed design has three components:
\subsubsection{Index Counter} Each column of \textbf{A} is indexed simultaneously with the corresponding row of \textbf{B}. This is coordinated with the help of an \textit{index counter} that counts from 0 to \textit{N}, incrementing each time the \textit{done} signal is asserted. Once the count reaches $N$, it asserts an \textit{output\_valid} signal, indicating the end of the matrix multiply computation.
% This is coordinated by a simple \textit{index counter} that counts from 0 to N, incrementing each time the enable signal (\textit{done\_left}) is high. 
% This counter will assert output\_ready when the count has reached N. One of these will exist in the architecture.
%
\subsubsection{Vector Generator} This component receives the index from the \textit{index counter} and outputs a corresponding $M$-dimensional column of \textbf{A} and $P$-dimensional row of \textbf{B}. Two \textit{vector generators} exist in the design - one each for \textbf{A} and \textbf{B}.
%
% \begin{figure}[t]
% \centering
% \includegraphics[width=0.6\columnwidth]{figures/tu_encoder.png}
% \caption{The Temporal-Unary Encoder}
% \label{fig:bd1}
% \end{figure}

\begin{figure}[t]
    \centering
    \begin{subfigure}[b]{0.6\columnwidth}
        \centering
        \includegraphics[width=\columnwidth]{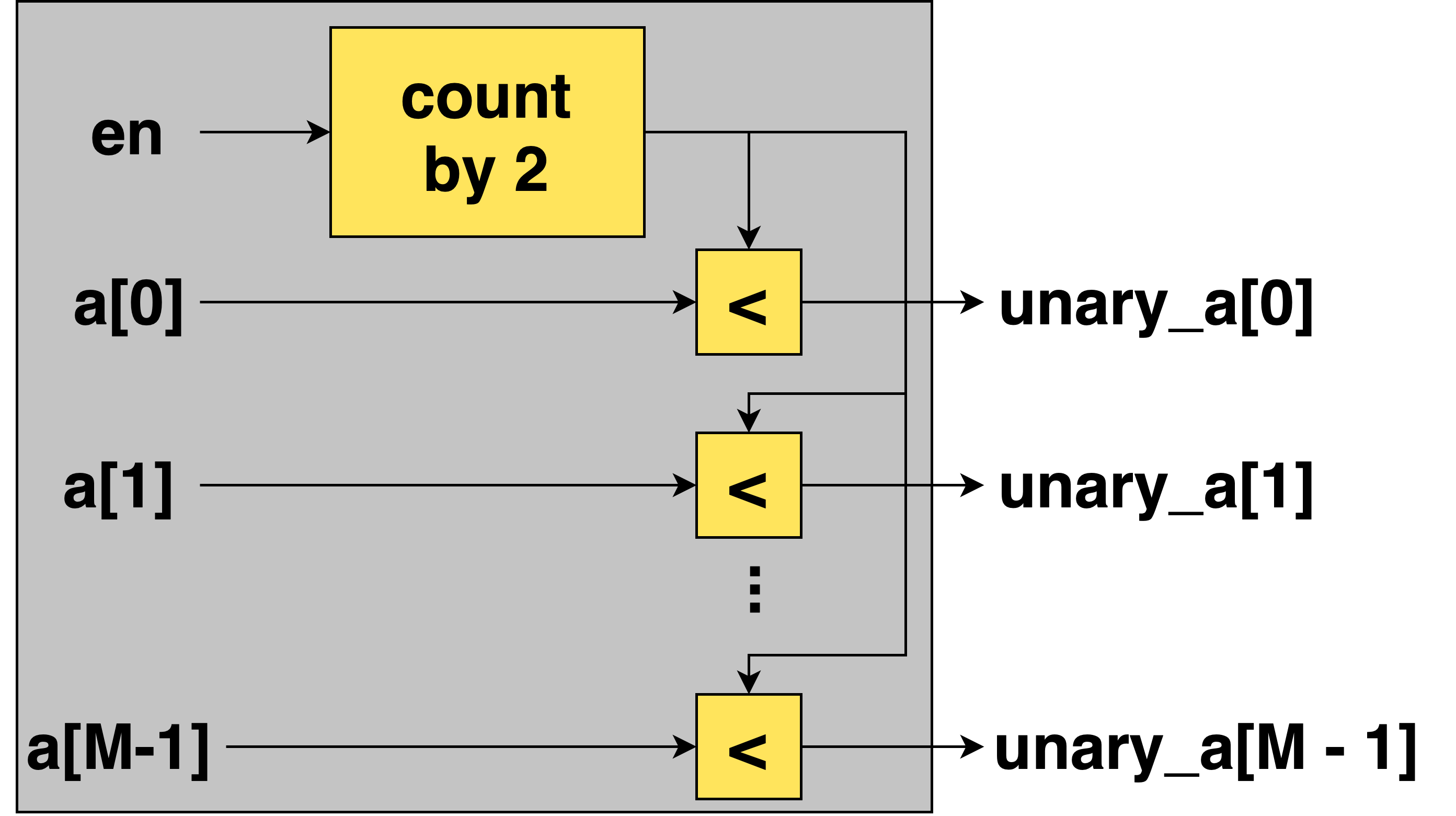}
        \caption{Temporal-Unary Encoder}
        \label{fig:bd1}
    \end{subfigure}%
    \begin{subfigure}[b]{0.19\textwidth}
        \centering
        \includegraphics[width=\columnwidth]{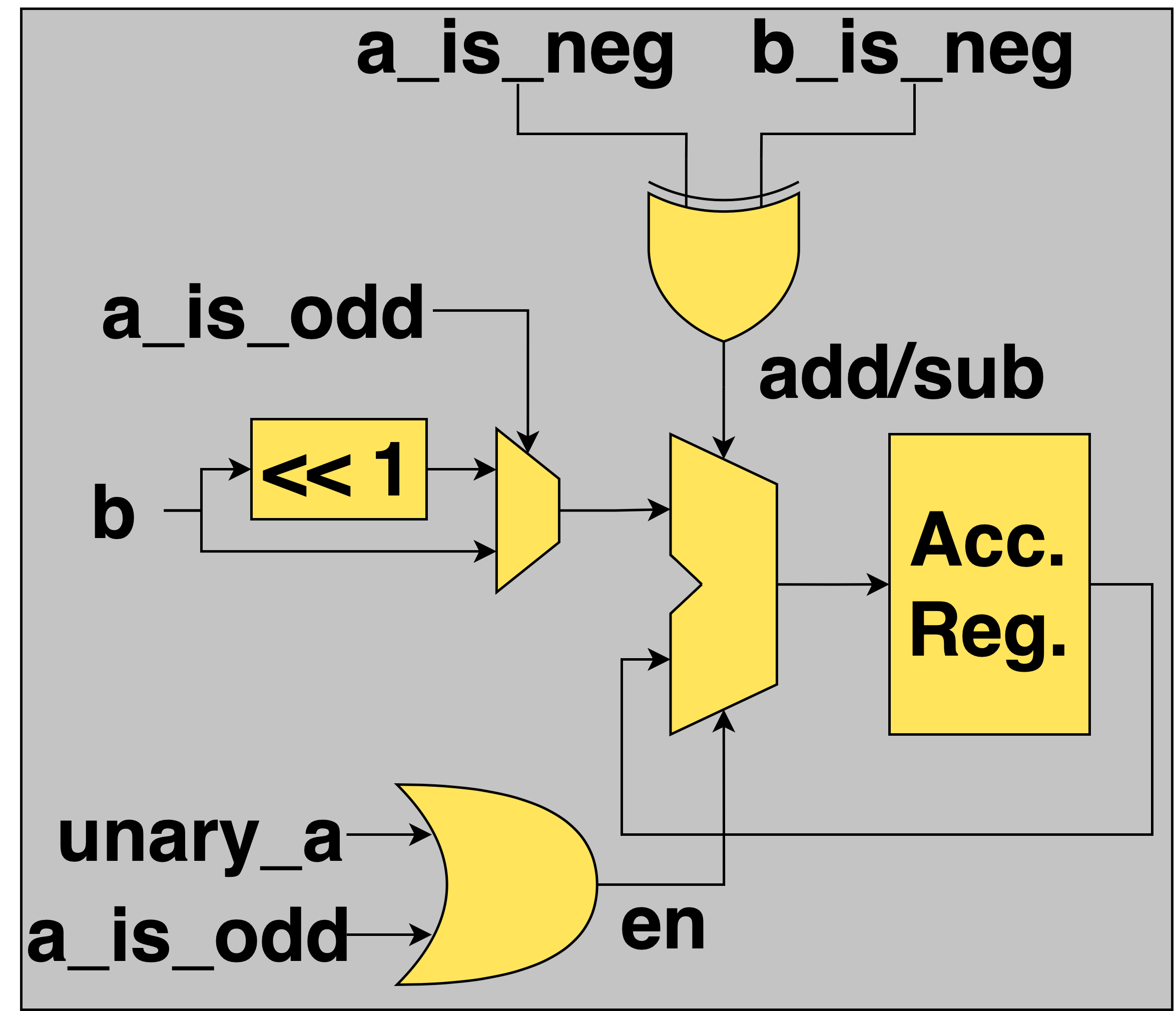}
        \caption{Processing Element}
        \label{fig:bd2}
    \end{subfigure}
    \caption{tubGEMM components}
\end{figure}

\subsubsection{Temporal-Unary Encoder}
\label{sec:tu_enc}
As shown in Fig. \ref{fig:bd1}, it consists of a single \textit{twos}-counter that counts up with increments of 2, and a set of $M$ comparators that convert the $M$ binary values within one column of \textbf{A} into $M$ unary signals. Each comparator asserts a high signal at the output for as long as the binary value is greater than the counter value ($\lfloor \frac{n}{2}\rfloor$ cycles for binary value $n$). Since each unary cycle counts up by two, odd values need to have a correction factor, which is facilitated by asserting \textit{a\_is\_odd} signal in the last cycle, sent to the PEs.
%(i.e., the very next cycle after the unary output is deasserted). This correction mechanism is explained further in the next subsection.

% \begin{figure}[t]
% \centering
% \includegraphics[width=0.6\columnwidth]{figures/bipolar_pe_mod.png}
% \caption{tubMAC Processing Element for Bipolar Computation}
% \label{fig:bd2}
% \end{figure}

\subsection{Multiply-Accumulate Processing Element}
\label{sec:mac}
As shown in Fig. \ref{fig:bd2}, each node in the PE array is a MAC unit that multiplies a unary signal with a binary signal and adds the resulting binary output to the previously stored binary value. We refer to each element within \textbf{A} and \textbf{B} matrices as \textbf{a} and \textbf{b}, respectively. Akin to T-MAC \cite{pan1t}, a sequential multiplier is used to accumulate the binary input for as many cycles as the unary input is asserted. We employ bipolar (signed) processing where both inputs \textbf{a} and \textbf{b} are assumed to be in twos-complement format (\textit{b}-bits wide), with the maximum magnitude being $2^{b-1}$. With the twos-unary encoding, a single multiplication operation can take up to $2^{b-2}$ cycles.
% (half the magnitude of the value encoded).

Additional mechanisms are employed to handle negative and odd inputs. \textit{a\_is\_odd} signal controls a multiplexer, which outputs \textbf{b} when it is enabled; 2*\textbf{b} otherwise. XOR of the most significant bits from \textbf{a} and \textbf{b} determines whether to add or subtract the output of the multiplexer to the accumulating register. The sequential multiplication is enabled through the OR of the unary input and \textit{a\_is\_odd} signal.

\section{Experimental Results}
\label{results_sec}
To perform fair comparisons with uGEMM \cite{ugemm}, tubGEMM is first synthesized using NanGate45 Open Cell Library using Synopsys Design Compiler. tubGEMM is configured to the same dimensions in \cite{ugemm} with 16x16 input matrices, 8-bit precision, and the clock speed set to 400 MHz.
% We further expand our results to include 7nm PPA using ASAP7 \cite{clark2016asap7} cell library with Cadence Genus for varying dimensions and value precision bit-widths. In both cases, Value Change Dump (VCD) from RTL simulation is sourced during synthesis to generate accurate post-synthesis power values.
Further, we extend the evaluation to get accurate 5nm PPA and energy results. The process design kit (PDK) and technology library used for evaluation is the industry-standard TSMC N5 (5nm) process node, with Synopsys design tools employed for simulation, synthesis, and power calculations. The tubGEMM RTL design is first created in SystemVerilog, with functional verification performed using Synopsys VCS. Consequently, lint check is performed on the SystemVerilog source file using Synopsys SpyGlass and then synthesis is performed to convert the RTL-level design into a gate-level netlist using Synopsys Design Compiler, sourcing TSMC N5 library files. Gate-level simulation is then performed for netlist verification and switching activity collection in the form of a SAIF dump. The SAIF dump is then sourced along with the netlist to perform accurate power calculations using Synopsys PrimeTime PX.

The experimental results presented in the following four subsections include: 1) Comparison of tubGEMM vs. other unary GEMM designs including uGEMM, in 45nm CMOS; 2) Design space evaluation of tubGEMM in TSMC N5 by scaling the matrix dimensions from 16x16 to 128x128, and bit precision from 2 bits to 8 bits; 3) Assessment of the inherent potential of tubGEMM to exploit dynamic data sparsity; and 4) Evaluation of tubGEMM for two real DNN workloads.

Sections \ref{sec:compare} and \ref{sec:5nm} derive latency and energy results based on the maximum possible cycle count for tubGEMM ($2^{b-1}$ for unipolar and $2^{b-2}$ for bipolar, with $b$-bit inputs) and is denoted as worst-case (WC). The (significantly better) average-case latency and energy values for real DNN workloads are presented in Sections \ref{sec:sparsity} and \ref{sec:dnnwl}.

\begin{figure}[t]
\centering
\includegraphics[width=0.95\columnwidth, height=14cm]{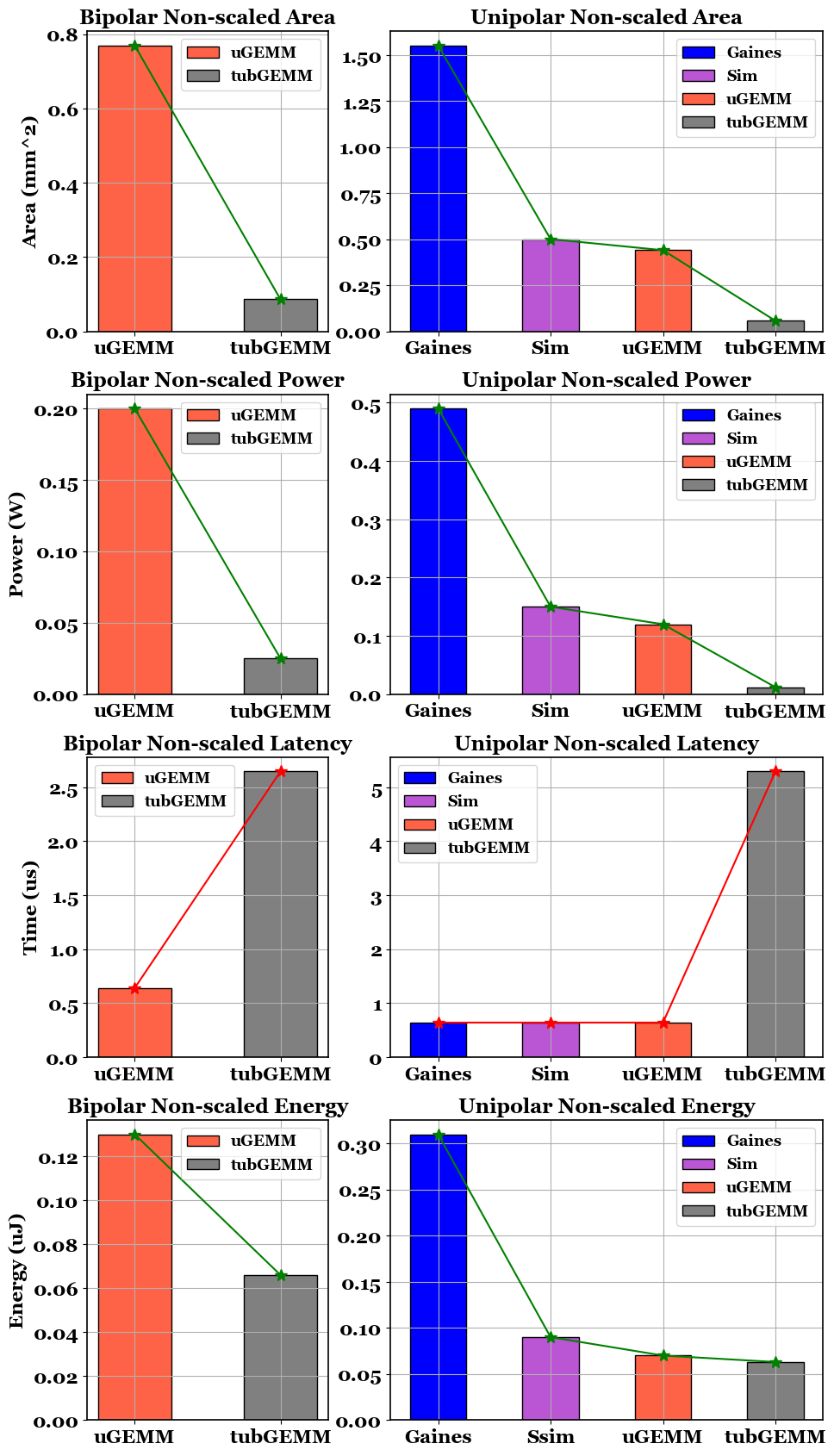}
\caption{uGEMM vs. tubGEMM 45nm post-synthesis WC PPA and energy values for bipolar and unipolar non-scaled GEMM}
\label{fig:baseline}
\end{figure}

\subsection{tubGEMM vs. Other Unary GEMM Designs}
\label{sec:compare}

 \begin{table}[t]
\centering
\caption{45nm Bipolar 8-bit Results for 16x16 matrices}
\scalebox{0.95}{
 \begin{tabular}{|c|c|c|c|c|c|c|} 
 \hline
 GEMM & Area & Power & Latency & Energy & Accuracy\\
 Hardware & (mm\textsuperscript{2}) & (W) & (us) & (uJ) & (\%)\\
 \hline
 \hline
  uGEMM & 0.77 & 0.20 & 0.64 & 0.13 & 61.37  \\
 \hline
 \textbf{tubGEMM} & \textbf{0.086} & \textbf{0.025} & \textbf{2.65 (WC)} & \textbf{0.066 (WC)} & \textbf{100}\\
 \hline
 \end{tabular}
 }
  \label{tab:bipolar}
\end{table}

\begin{table}[t]
\centering
\caption{45nm Unipolar 8-bit Results for 16x16 matrices}
\scalebox{0.95}{
 \begin{tabular}{|c|c|c|c|c|c|c|} 
 \hline
 GEMM & Area & Power & Latency & Energy & Accuracy\\
 Hardware & (mm\textsuperscript{2}) & (W) & (us) & (uJ) & (\%)\\
 \hline
 \hline
  Gaines & 1.55 & 0.49 & 0.64 & 0.31 & 70.45  \\
 \hline
  Sim & 0.50 & 0.15 & 0.64 & 0.09 & 70.93  \\
 \hline
 uGEMM & 0.44 & 0.12 & 0.64 & 0.07 & 100  \\
 \hline
 \textbf{tubGEMM} & \textbf{0.057} & \textbf{0.012} & \textbf{5.29 (WC)} & \textbf{0.063 (WC)} & \textbf{100}\\
 \hline
 \end{tabular}
 }
  \label{tab:unipolar}
\end{table}

We compare against Gaines \cite{gaines1967stochastic}, Sim \cite{sim2017new} and uGEMM results for bipolar and unipolar non-scaled implementations as provided in uGEMM \cite{ugemm}.
We illustrate tubGEMM's results for both bipolar (signed) and unipolar (unsigned) GEMM operations in Fig. \ref{fig:baseline}, and the numbers are provided in Tables \ref{tab:bipolar} and \ref{tab:unipolar}, respectively, along with the computation accuracy values. As reported in Table \ref{tab:bipolar}, bipolar tubGEMM outperforms bipolar uGEMM in all hardware metrics by 9x on area (89\% less), 8x on power (87\% less), and 2x on energy (50\% less), while increasing latency by 4.1x. A key highlight here is that on top of being much more hardware efficient, tubGEMM achieves ideal accuracy as compared to uGEMM's reported accuracy of 61.37\% for temporal coding.
 
Among the unipolar non-scaled designs (Table \ref{tab:unipolar}), tubGEMM again provides better metrics except latency, which is double the bipolar non-scaled latency number. Sim \cite{sim2017new}, Gaines \cite{gaines1967stochastic}, and uGEMM \cite{ugemm} require 8.3x less latency compared to tubGEMM. However, the high latency is offset by the other metrics. To detail further, compared to uGEMM which is the best performing out of the existing unary approaches, tubGEMM is 7.7x, 10x, and 1.1x more optimized for area, power, and energy consumption, respectively. Additionally, it maintains the ideal accuracy results demonstrated by uGEMM. These results show that tubGEMM can be a promising candidate for edge-native DLAs because of its extremely hardware-efficient implementation and low energy consumption. We address its one shortcoming of longer latency in Section \ref{sec:sparsity}.

\textbf{Key Takeaway}: Compared to current state-of-the-art uGEMM, tubGEMM can achieve significant improvements in area, power, and energy, even with its worst-case latency.

\begin{table}[t]
\centering
\caption{TSMC N5 post-synthesis tubGEMM PPA and energy values for various bit-widths and input matrix sizes}
\scalebox{0.95}{
 \begin{tabular}{|c|c|c|c|c|c|c|} 
 \hline
 PE & Bit & Area & Power & WC-Latency & WC-Energy\\
 Array & -width& (um\textsuperscript{2}) & (mW) & (us) & (nJ)\\
 \hline
 \hline
  16 x 16 & 8-bit & 2777.71 & 3.75 & 2.65 & 9.93\\
 \hline
     & 4-bit & 1275.12 & 1.66 & 0.25 & 0.42\\
 \hline
     & 2-bit & 692.50 & 0.91 & 0.13 & 0.12  \\
 \hline
  32 x 32 & 8-bit & 12560.52 & 21.80 & 5.30 & 115.55\\
 \hline
     & 4-bit & 5270.51 & 8.41 & 0.50 & 4.21\\
 \hline
     & 2-bit & 2804.12 & 3.19 & 0.26 & 0.83\\
 \hline
 64 x 64 & 8-bit & 50384.67 & 85.13 & 10.60 & 902.33\\
 \hline
    & 4-bit & 22507.20 & 39.07 & 1.00 & 39.07\\
 \hline
    & 2-bit & 12108.43 & 16.10 & 0.52 & 8.37\\
 \hline
 128 x 128 & 8-bit & 221689.65 & 417.72 & 21.20 & 8855.72 \\
 \hline
        & 4-bit & 99030.84 & 209.91 & 2.00 & 419.82\\
 \hline
        & 2-bit & 53765.53 & 96.45 & 1.04 & 100.31\\

 \hline

 \end{tabular}
 }
  \label{tab:5nm}
\end{table}

\subsection{5nm CMOS Scaling of tubGEMM}
\label{sec:5nm}
\begin{figure}[t]
\centering
\includegraphics[width=0.9\columnwidth, height=14.2cm]{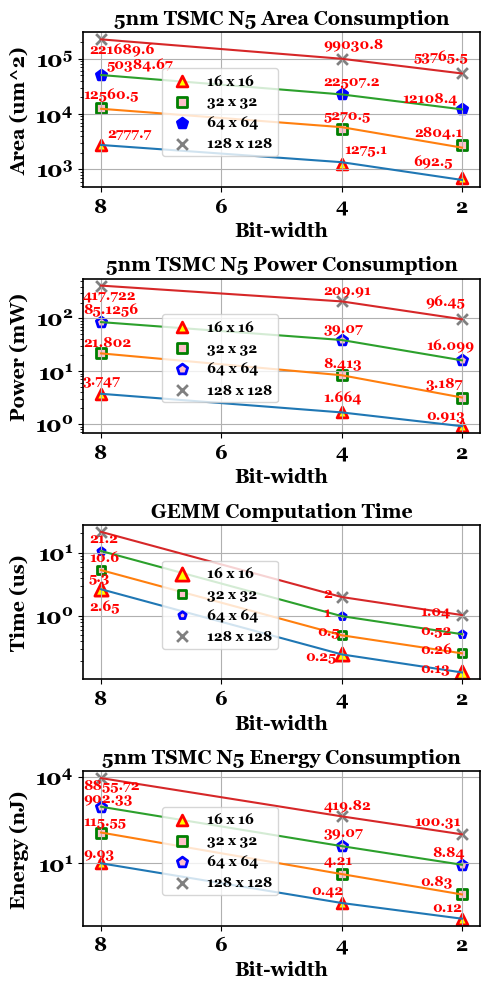}
\caption{tubGEMM TSMC N5 PPA and energy scaling, for input matrix dimensions of 16x16, 32x32, 64x64, and 128x128, and matrix element value bit-widths of 8 bits, 4 bits, and 2 bits}
\label{fig:5nm}
\end{figure}

We depict tubGEMM's hardware complexity in 5nm CMOS as we scale input matrix (or MAC array) dimensions: 16x16, 32x32, 64x64, and 128x128, and bit precision: 8-bit, 4-bit, and 2-bit in Fig. \ref{fig:5nm} (y-axis is in log scale), with its corresponding 
%Results for 128x128 are not plotted for brevity; it follows the same trend as other configurations.
% The y-axis of the plots is in log-scale to better visualize the scaling trends.
post-synthesis numbers in Table \ref{tab:5nm}. As seen in the latency subplot in Fig. \ref{fig:5nm}, latency scales linearly with matrix size and has a noticeable speedup upon transitioning from 8-bit to 4-bit precision due to an exponential decrease in cycle count. 
% As observed in the latency subplot, the latency numbers are agnostic to the input dimension size scaling but have a noticeable speedup upon transitioning from an 8-bit resolution to a 4-bit resolution owing to an \textit{11x decrease in clock cycle count}.

\subsubsection{Hardware Complexity Scaling with Increasing Input-Matrix Dimensions}
The area, power, and energy consumption values scale nearly quadratically with an increase in the input dimension. This is equivalent to a linear increase with the number of PEs. With each doubling of one dimension of the input matrices, or MAC arrays, the tubGEMM design scales approximately 4.3x, 4.9x, and 9.8x for the area, power, and energy consumption on average, respectively. Since the clock frequency is set at 400 MHz, cell leakage power scales almost precisely with the area utilized.

\subsubsection{Hardware Complexity Scaling with Decreasing Element Value Bit-width}
Here, the input dimension size is set constant, and the scaling trend for decreasing bit-width precision is analyzed. On average, lowering from 8-bit to 4-bit compute decreases area, power, and energy by 2.3x, 2.4x, and 23.9x, respectively. 
% As explained earlier,
The considerable reduction in energy from 8-bit to 4-bit is due to an 11x decrease in latency. Further, transitioning from 4-bit to 2-bit compute results in 1.8x, 2.2x, and 4.4x reduction in area, power, and energy, respectively.

\textbf{Key Takeaway:} The tubGEMM design for 8-bit 128x128 matrices (same as Google TPU v3) consumes only 0.22 mm\textsuperscript{2} and 418 mW. Lowering the precision reduces the energy from 8,856 nJ (8-bit) to 420 nJ (4-bit) and 100 nJ (2-bit); this bodes well with the projected trend toward 4-bit and 2-bit DLAs \cite{wang20188}.

% \subsection{Gate-level Analysis}
% As observed before, since the matrices are composed of Two’s Complement numbers, it will take 2\^(B-1) cycles to represent the largest magnitude number. Using the counting by 2s optimization, it is possible to reduce it to 2\^(B-2) + 1 cycles. Since this final architecture still requires looping over all the rows of the input matrix, you must multiply this number of cycles by the number of rows N. Accounting for some additional overhead, the worst-case latency of this design is N*(2\^(B-2) + 2) + 2 cycles. This architecture ended up being a very good balance of area and latency resulting in a respectable energy-efficiency.

% Below is a rough calculation of the area required for this design (note the red text is additional area exclusive to scaled matrix multiplication):

\subsection{Dynamic Sparsity Evaluation}
\label{sec:sparsity}
% \begin{figure}[t]
% \centering
% \includegraphics[width=\columnwidth]{tops_sp.png}
% \caption{Energy-efficiency (TOPS/W) Scaling with Sparsity}
% \label{fig:sparsity}
% \end{figure}
% 

The tubGEMM latency values illustrated above reflect the worst-case scenario, assuming each multiply involves the largest possible absolute value (127 for bipolar and 255 for unipolar with 8-bit values), thereby taking a proportional number of cycles to finish one compute. However, as a result of less frequently occurring large values and frequently occurring zeros in typical DL compute, tubGEMM can naturally leverage such bit-level and word-level value sparsities to significantly reduce the effective compute latency. Here, word-level sparsity refers to zero values in typical weights and activations, which has been widely explored in literature. On the other hand, bit-level sparsity here refers to a small number of ones followed by many zeros in the unary bitstream. As the value of a temporal-unary encoded signal is equal to the number of ones in the bitstream, bit-level sparsity directly results from small magnitude values. Bit-level sparsity subsumes word-level sparsity in the extreme case when all bits in the unary bitstream are zeros. Hence, we focus on bit sparsity here.

\begin{figure}[!t]
\centering
\includegraphics[width=0.85\columnwidth]{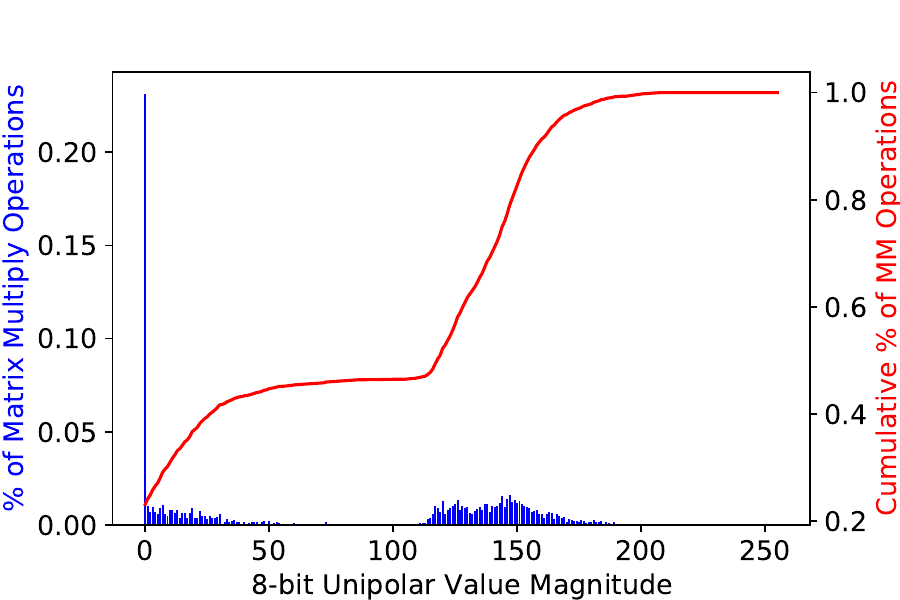}
\caption{Bit Sparsity: Percentage of matrix multiply operations that involve the corresponding X-axis values as the maximum magnitude during inference of INT8 unipolar Quantized MobileNetv2. Right Y-axis plots the cumulative percentage of matrix multiply operations with maximum values less than or equal to the corresponding X-axis value.}
\label{fig:percentmv2}
\end{figure}

% \subsubsection{Bit-Sparsity Impact on Latency and Energy}

% Here, we investigate the bit sparsity present in a typical edge AI/DL scenario and assess its impact on the tubGEMM compute latency and energy. 
To profile bit sparsity, we perform inference on a representative DNN workload, namely, pre-trained unipolar INT8 quantized MobileNetv2, and keep track of the maximum values in every feature map within each convolutional layer. The underlying assumption is that each feature map computation can be packaged as a matrix multiply operation, whose latency is bounded by the maximum magnitude value in that feature map. Inference is performed using PyTorch, and the pre-trained model is taken from Torchvision library. During inference, we calculate the number of times every value between 0 and 255 (maximum magnitude for 8-bit unipolar compute) manifests as the maximum value within a feature map. 
% This frequency of occurrence of maximum values per inference averaged across 1000 random input images is plotted in Fig. \ref{fig:freqmv2}. Note that the Y-axis is in log scale. It can be seen that `0' occurs most commonly (about 700 times) as the maximum value within a tubGEMM compute and the value 255 rarely occurs. In contrast, all the other values occur less than 50 times. 
This occurrence of maximum values derived as a percentage of the total number of matrix multiply operations per inference is averaged across 1000 randomly selected input images from ImageNet and plotted in blue in Fig. \ref{fig:percentmv2} along with the cumulative percentage in red. Fig. \ref{fig:percentmv2} clearly illustrates that close to 25\% of the operations involve only zeros and thereby take zero cycles; any other value occurs less than 2\% of the times (255 rarely occurs). Cumulatively, 90\% of tubGEMM operations in MobileNetv2 involve values smaller than 150.

\begin{table}[t]
\centering
\caption{45nm average-Case Latency, Energy and Energy-Delay Product for Unipolar Non-Scaled 16x16 tubGEMM based on Bit-Sparsity in INT8 Quantized MobileNetv2}
\scalebox{1}{
 \begin{tabular}{|c|c|c|c|} 
 \hline
 16x16 GEMM & Latency & Energy & Energy-Delay\\
 Hardware & (us) & (uJ) & Product (uJ-us) \\
 \hline
 \hline
  uGEMM & 0.64 & 0.070 & 0.045 \\
 \hline
  tubGEMM (worst-case) & 5.29 & 0.063 & 0.333 \\
 \hline
 \textbf{tubGEMM (average-case)} & \textbf{1.72} & \textbf{0.021} & \textbf{0.036} \\
 \hline
 \end{tabular}
 }
  \label{tab:avglat}
\end{table}

From Fig. \ref{fig:percentmv2}, the expected maximum feature map value can be calculated using the area under the blue curve, which gives 82. Based on this data, the average-case tubGEMM latency for MobileNetv2 comes out to be 1.72 us (3x reduction from 5.29 us), leading to a 3x reduction in energy as can be seen from Table \ref{tab:avglat}. It can also be seen that the Energy-Delay Product (EDP) of tubGEMM reduces even below that of uGEMM (1.25x improvement over uGEMM and 9.25x improvement over worst-case tubGEMM). This demonstrates the efficacy of tubGEMM in typical edge DL scenarios where much of the exponential overhead in latency is hidden due to the sparse nature of input data values (both weights and activations).

\textbf{Key Takeaway:} tubGEMM can implicitly exploit sparsity in data (both bit sparsity and word sparsity) to further improve all three metrics: power, latency, and energy. This dynamic form of data-dependent run-time optimization is built into the tubGEMM design. As a result, dynamic power is reduced, and the effective latency is much smaller than the worst-case latency, leading to a significant reduction in energy and EDP.

\subsection{DNN Workload Analysis}
\label{sec:dnnwl}
In this section, we build on the bit sparsity analysis from Section \ref{sec:sparsity} and perform RTL hardware benchmarking with two real DL workloads, specifically, MobileNetv2 and ResNet-50, which are both INT8 quantized. Representative test vectors are generated from weights and activations in both these workloads for tubGEMM with 64x64 PE array and 8-bit precision, and are used to produce switching activity files for logic synthesis. The average compute latency per GEMM operation is derived through RTL functional simulation and then combined with post-synthesis power consumption to calculate workload-specific energy consumption and energy-delay product (EDP), which are presented in Table \ref{tab:dnnwl}. Note that 64x64 tubGEMM PE array size is used for evaluation (Table \ref{tab:dnnwl}) since it is a realistic configuration used in Edge Tensor Processing Units (TPUs) \cite{8701189}, as opposed to 16x16 array size used in Table \ref{tab:avglat} to match with the baseline uGEMM.

\begin{table}[t]
\centering
\caption{TSMC N5 Post-synthesis Power, Latency, Energy and Energy-Delay Product for representative DNN workloads, MobileNetv2 and ResNet-50, on 64x64 tubGEMM}
\scalebox{1}{
 \begin{tabular}{|c|c|c|c|c|} 
 \hline
 Workload & Power & Latency & Energy & Energy-Delay\\
 Type & (mW) & (us) & (uJ) & Product (uJ-us) \\
 \hline
 \hline
 Random (Table \ref{tab:5nm}) & 85.13 & 10.60 & 0.90 & 9.54\\
 \hline
 \hline
 MobileNetv2 & 49.43 & 5.54 & 0.27 & 1.52\\
 \hline
 ResNet-50 & 56.35 & 4.66 & 0.26 & 1.22\\
 \hline
 \end{tabular}
 }
  \label{tab:dnnwl}
\end{table}

The first entry in Table \ref{tab:dnnwl} corresponds to the 64x64 8-bit values from Table \ref{tab:5nm}, which are generated using uniform random test vectors. As can be seen from Table \ref{tab:dnnwl}, by leveraging dynamic activation and weight sparsity present within real MobileNetv2 and ResNet-50 workloads, tubGEMM is able to reduce power, latency, energy and EDP by 1.7x, 1.9x, 3.2x and 6.3x respectively for MobileNetv2, and 1.5x, 2.3x, 3.4x and 7.7x respectively for ResNet-50.
In Sec \ref{sec:compare}, we illustrated tubGEMM's superiority over existing unary designs, including uGEMM, while still considering worst-case latency.
% The analysis presented here demonstrates that real workloads enable tubGEMM to further enhance this superiority significantly.
This clearly demonstrates that tubGEMM, in spite of its long worst-case latency, can run real DNN workloads with much lower latency and very high energy efficiency.

\textbf{Key Takeaway:} Compared to worst-case latency, dynamic power and energy consumption incurred by tubGEMM are significantly lower for typical real DNN workloads due to high word-level and bit-level sparsity in weights and activations.

\section{Concluding Remarks}
%TODO: Mention about on-chip SRAM elimination like usystolic

%\subsection{Concluding Remarks and Future Extensions}

We present a novel GEMM design, tubGEMM, that operates on temporal-unary encoded data, performs exact computation, and significantly reduces area, power, and energy (by 89\%, 87\%, and 50\% respectively) relative to state-of-the-art 
% binary MXUs as well as the leading 
unary GEMM design, uGEMM. We implement tubGEMM in 45nm as well as 5nm CMOS and show very promising PPA and energy results for scaling from 16x16 to 128x128 matrix multiply and for precision ranging from 8 bits down to 2 bits. 
% Our results illustrate that reducing the precision from 8 bits to 4 bits and 2 bits improves energy by almost 13x and 37x respectively.
% , with the energy consumption being ranging from a few tens or hundreds of nJ to less a nJ. 
% We believe, tubGEMM, which can seamlessly ingest binary data (and convert to \textit{tub} internally), can be widely adopted for low-precision DLAs for edge devices.
% 
tubGEMM can efficiently and naturally exploit data sparsity in DL computations
% for both training and inference. It can automatically adapt to and exploit sparsity of data value
at both word and bit levels, significantly reducing latency as well as dynamic power consumption. Typical sparsity in DNNs can potentially achieve 3x and 8x reduction in energy and EDP, respectively. This can facilitate energy-efficient online continuous learning for edge devices. We believe, tubGEMM, which can seamlessly ingest binary data (and convert to \textit{tub} internally), can be widely adopted as a matrix multiply unit within low-precision edge-native DLAs. Our future work will extend the hardware design of the matrix multiply unit, focus of this paper, to system-level DLA design incorporating memory subsystem in TSMC 5nm process node.
% Furthermore, tubGEMM can potentially reduce the memory bandwidth requirements due to its temporal-unary encoding and bit-serial processing.

%%
%% The next two lines define the bibliography style to be used, and
%% the bibliography file.
\bibliographystyle{IEEEtranS}
\bibliography{refs}

\end{document}